\documentclass[10pt,draft]{dis03}
\usepackage{epsf,amsmath}

\begin{document}
\titlepage
\begin{flushright}
\hfill IPPP/03/43 \\
\hfill DCPT/03/86\\
\hfill Cavendish-HEP-2003/13\\
\hfill 21 July 2003 \\
\end{flushright}

\vspace*{0.5cm}

\begin{center}
{ \bf MRST partons and uncertainties
}

\vspace*{0.5cm}

{A.D.~Martin$^1$, R.G.~Roberts$^1$, W.J.~Stirling$^1$ and R.S.~Thorne$^2$ \\[1.2ex]
$^1$IPPP, Durham, DH1 3LE, UK\\
$^2$Cavendish Laboratory, Cambridge, CB3 0HE, UK \\[1.2ex]
E-mail: a.d.martin@durham.ac.uk }

\end{center}

\begin{abstract}
\noindent We discuss uncertainties in the extraction of parton distributions from global analyses of DIS and
related data. We present {\em conservative} sets of partons, at both NLO and NNLO, which are stable to $x,Q^2,W^2$
cuts on the data. We give the corresponding values of $\alpha_S(M_Z^2)$ and the cross sections for $W$ production
at the Tevatron.
\end{abstract}

The parton distributions of the proton, which are currently determined from global analyses of a wide range of DIS
and related hard scattering data, are subject to many sources of uncertainty.  There are uncertainties due to
(i)~the experimental errors on the data that are fitted in the global analysis, (ii)~the choice of data cuts
($W_{\rm cut}, x_{\rm cut}, Q^2_{\rm cut}$), defined such that data with values of $W$, $x$ or $Q^2$ below the cut
are excluded from the global fit, (iii)~the truncation of the DGLAP perturbation expansion, (iv)~specific
theoretical effects, such as $\ln\,1/x$, $\ln(1-x)$, absorptive and higher-twist corrections, and (v)~input
assumptions, such as isospin invariance, the choice of parameterization, heavy target corrections and the form of
the strange quark sea.

So far, attention has been focussed on the uncertainties arising from the experimental errors; see
Refs.~\cite{CTEQ,MRST2002} for estimates based on global NLO analyses. However, Ref.~\cite{MRST2003} concentrates
on the remaining uncertainties, (ii)--(v); here we present some results from this forthcoming paper. In principle,
if the DGLAP formalism is valid and the various data sets are compatible, then changing the data that are included
in the global analysis should not move the predictions outside the error bands. In practice this is not the case.
Consider, for instance, the effect of different choices of $x_{\rm cut}$ on the data that are fitted. Table~1
shows the values of $\chi^2$ for NLO global analyses performed for different values of $x_{\rm cut}$, together
with the number of data points fitted. Each column represents the $\chi^2$ values corresponding to a fit performed
with a different choice of the cut in $x$.


\begin{table}[!thb]\begin{center}
\begin{tabular}{|c|c|c|c|c|c|c|} \hline
$x_{\rm cut}:$ & 0 & 0.0002 & 0.001 & 0.0025 & 0.005 & 0.01 \\
\hline \# data points & 2097 & 2050 & 1961 & 1898 & 1826 & 1762 \\
\hline $\chi^2(x>0)$ & 2267 &&&&& \\
$\chi^2(x>0.0002)$ & 2212 & 2203 &&&& \\
$\chi^2(x>0.001)$ & 2134 & 2128 & 2119 &&& \\
$\chi^2(x>0.0025)$ & 2069 & 2064 & 2055 & 2040 && \\
$\chi^2(x>0.005)$ & 2024 & 2019 & 2012 & 1993 & 1973 & \\
$\chi^2(x>0.01)$ & 1965 & 1961 & 1953 & 1934 & 1917 & 1916 \\
\hline $\Delta_i^{i+1}$ & \multicolumn{6}{| l |}{$\qquad\ \  0.19
\quad\ \ \,  0.10 \quad\ \ \;  0.24 \quad\ \ \  0.28 \quad\ \ \ \,  0.02\quad\;  $} \\
\hline
\end{tabular} \caption{\label{tab:t1}
The measure of stability, $\Delta_i^{i+1}$, to changing the choice of $x_{\rm cut}$.}
\end{center}
\end{table}

To obtain a measure of the stability of the analysis to changes in the choice of $x_{\rm cut}$, we compare fits in
adjacent columns, that is with $(x_{\rm cut})_{i+1}$ and $(x_{\rm cut})_i$. In particular, it is informative to
compare the contributions to their respective $\chi^2$ values from the subset of data with $x>(x_{\rm
cut})_{i+1}$. If stability were achieved, then we would expect the difference $\Delta \chi^2$ between these two
$\chi^2$ contributions to be very small. We stress that these two $\chi^2$ contributions describe the quality of
the two fits to the {\em same} subset of the data. Thus a measure, $\Delta_i^{i+1}$, of the stability of the
analysis is $\Delta\chi^2$ divided by the number of data points omitted when going from the fit with $(x_{\rm
cut})_i$ to the fit with $(x_{\rm cut})_{i+1}$. For example, if we raise the $x_{\rm cut}$ from 0.001 to 0.0025
then $\Delta\chi^2 = 2055 - 2040$ for the data with $x>0.0025$, and the number of data points omitted is $1961 -
1898 = 63$. Thus the measure $\Delta_{0.001}^{0.0025} = 15/63 = 0.24$, as shown in the last row of Table~1.

Inspection of the values of $\Delta_i^{i+1}$ shows a significant improvement in the quality of the fit each time
$x_{\rm cut}$ is raised until the final step when $x_{\rm cut}$ is increased from 0.005 to 0.01, when we see that
there is no further improvement at all. In fact, raising $x_{\rm cut}$ from 0.01 to 0.02 confirms this stability.
Hence we conclude that $x\simeq 0.005$ is a safe choice of $x_{\rm cut}$.

After similar studies of the effect of varying $Q^2_{\rm cut}$ and $W^2_{\rm cut}$, and combinations of these
cuts, we find a restricted domain ($x_{\rm cut} > 0.005$, $Q^2_{\rm cut}>10~{\rm GeV}^2$, $W^2 > 15~{\rm GeV}^2$)
in which the obtained NLO parton distributions are stable to further cuts. We denote this conservative parton set
by MRST03(cons). Their behaviour relative to MRST02~\cite{MRST2002} is shown in Fig.~1.

\begin{figure}[!thb]
\vspace*{1cm}
\begin{center}
\includegraphics{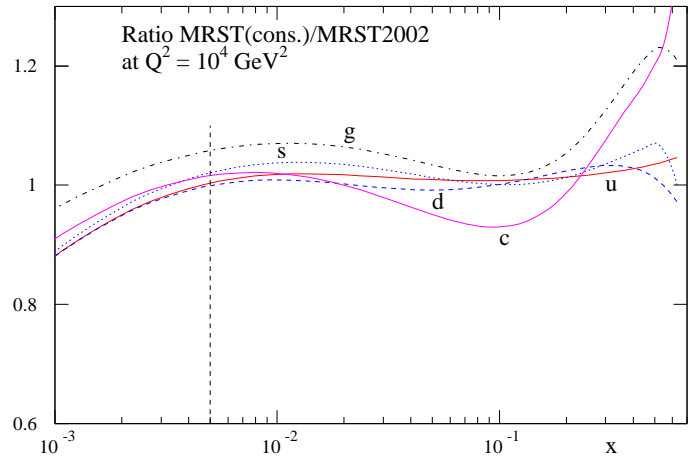} \vspace*{5.2cm} \caption{The {\em
conservative} partons compared to MRST2002~\cite{MRST2002}. \label{fig:f1} }
\end{center}
\vspace*{-0.5cm}
\end{figure}

 Clearly there are uncertainties associated with truncation of the DGLAP evolution at NLO. Now,
the DIS coefficient functions are known at NNLO~\cite{CF}. Also valuable, almost complete, information has been
obtained about the \linebreak NNLO splitting functions~\cite{SF}; indeed a range of compact analytic functions
exist that are all compatible with this information~\cite{VV12}. We therefore study the effect of data cuts at
NNLO. In going from NLO to NNLO the stable domain ($x_{\rm cut}>0.005$, $Q^2_{\rm cut}>7~{\rm GeV}^2$, $W^2_{\rm
cut}>15~{\rm GeV}^2$) has not increased as much as we might have hoped. Nevertheless there are advantages in going
to NNLO~\cite{MRST2003}.

To illustrate the value of having `conservative' parton sets\footnote{These sets are available at
http://durpdg.dur.ac.uk/hepdata/mrs.html} at both NLO and NNLO we consider, as important examples, the
determination of $\alpha_S(M_Z^2)$ from DIS data and the prediction of the cross section for $W$ production,
$\sigma_W$, at the Tevatron. The values of $\alpha_S$ found in the NLO and NNLO global fits that produced the {\em
conservative} sets of partons are given in Table~\ref{tab:t2} corresponding to the MRST03 entry, together with
other recent determinations from DIS fits. The quoted errors reflect the tolerance $\Delta\chi^2$ used in the
various analyses. Remarkably, the determinations of $\alpha_S(M_Z^2)$ have converged approximately to a common
value, even though they are based on different selections of the data.

\begin{table}[!htb]\begin{center}\begin{tabular}{l c l}
\hline\hline
\rule[-2ex]{0ex}{5ex} &$\Delta\chi^2$&$\alpha_S(M_Z^2)\ \pm{\rm expt}
\pm{\rm theory}\pm{\rm model}$\\
\hline
\rule[-2ex]{0ex}{5ex}\fbox{NLO}&&\cr CTEQ6&  100 & $0.1165\pm0.0065$ \cr ZEUS & 50 & $0.1166\pm0.0049
~~~~~~~~~~~~\pm0.0018$ \cr MRST03 & 5 & $0.1165\pm0.002~\,\pm0.003$ \cr H1 & 1  & $0.115~\,\pm0.0017\pm0.005 ~~
^{+0.0009}_{-0.0005}$ \cr Alekhin & 1 & $0.1171\pm0.0015\pm0.0033$ \cr
\hline
\rule[-2ex]{0ex}{5ex}\fbox{NNLO} &&\cr MRST03 & 5 & $0.1153\pm0.002\pm0.003$ \cr Alekhin & 1 &
$0.1143\pm0.0014\pm0.0009$ \cr
\hline\hline
\end{tabular}
\caption{\label{tab:t2} The values of $\alpha_S(M_Z^2)$ found in NLO and NNLO fits to DIS data. The experimental
errors quoted correspond to an increase $\Delta\chi^2$ from the best fit value of $\chi^2$. CTEQ6~\cite{CTEQ6} and
MRST03 are global fits. H1\cite{H1Krakow} fit only a subset of the $F_2^{ep}$ data, while Alekhin~\cite{alekhin03}
also includes $F_2^{ed}$ and ZEUS~\cite{ZEUSfit} in addition include $xF_3^\nu$ data.}
\end{center}\end{table}

\begin{figure}[!thb]
\vspace*{6.0cm}
\begin{center}
\includegraphics{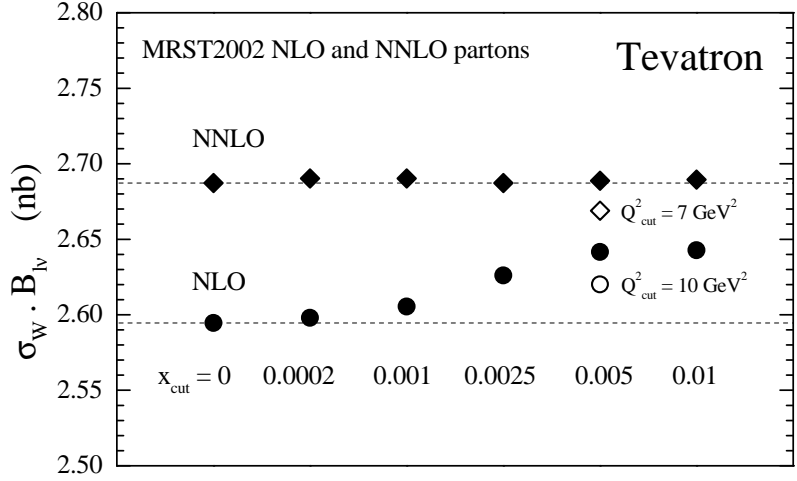}
\caption[*]{\label{fig:f2} Predictions for $W$ production at the Tevatron, $\sqrt s = 1.96$~TeV, for various
values of $x_{\rm cut}$, and for the conservative sets of partons (shown by the open symbols).}
\end{center}
\end{figure}

Fig.~\ref{fig:f2} shows values of the $W$ production cross section (times the leptonic branching ratio $B_{l\nu} =
0.1068$) at the Tevatron energy $\sqrt s = 1.96$~TeV, obtained from NLO and NNLO global fits of data, subject to
various values of $x_{\rm cut}$ and $Q^2_{\rm cut}$.
We see that the NNLO predictions are much more stable to variations of $x_{\rm cut}$. Note that at NNLO the
conservative parton set predicts a small decrease of 0.7\% relative to the NNLO default prediction (with $x_{\rm
cut} = 0$, $Q^2_{\rm cut} = 2~{\rm GeV}^2$), while at NLO there is an increase. The conservative partons thus show
greater convergence with increased perturbative order, than the default predictions. The NNLO conservative value
$B_{l\nu}\sigma_W = 2.67$~nb, with an expected total theoretical and experimental uncertainty of about 2\%, may
act as a very good luminosity monitor at the Tevatron.

\end{document}